\documentclass[aps, pra, twocolumn, amsfonts, amsmath, amssymb, superscriptaddress, floatfix]{revtex4}

\usepackage{graphicx}
\usepackage{dcolumn}
\usepackage{bm}
\usepackage{epsfig}
\usepackage{hyperref}
\usepackage{braket}
\usepackage{color}

\begin{document}

\title{Spin-Dipole Oscillation and Polarizability of a Binary Bose-Einstein Condensate near the Miscible-Immiscible Phase Transition}

\author{Tom Bienaim\'e}
\email[]{tom.bienaime@unitn.it}
\affiliation{INO-CNR BEC Center and Dipartimento di Fisica, Universit\`a di Trento, 38123 Povo, Italy}

\author{Eleonora Fava}
\affiliation{INO-CNR BEC Center and Dipartimento di Fisica, Universit\`a di Trento, 38123 Povo, Italy}

\author{Giacomo Colzi}
\affiliation{INO-CNR BEC Center and Dipartimento di Fisica, Universit\`a di Trento, 38123 Povo, Italy}
\affiliation{Trento Institute for Fundamental Physics and Applications, INFN, 38123 Povo, Italy}

\author{Carmelo Mordini}
\affiliation{INO-CNR BEC Center and Dipartimento di Fisica, Universit\`a di Trento, 38123 Povo, Italy}

\author{Simone Serafini}
\affiliation{INO-CNR BEC Center and Dipartimento di Fisica, Universit\`a di Trento, 38123 Povo, Italy}

\author{Chunlei Qu}
\affiliation{INO-CNR BEC Center and Dipartimento di Fisica, Universit\`a di Trento, 38123 Povo, Italy}

\author{Sandro Stringari}
\affiliation{INO-CNR BEC Center and Dipartimento di Fisica, Universit\`a di Trento, 38123 Povo, Italy}
\affiliation{Trento Institute for Fundamental Physics and Applications, INFN, 38123 Povo, Italy}

\author{Giacomo Lamporesi}
\affiliation{INO-CNR BEC Center and Dipartimento di Fisica, Universit\`a di Trento, 38123 Povo, Italy}
\affiliation{Trento Institute for Fundamental Physics and Applications, INFN, 38123 Povo, Italy}

\author{Gabriele Ferrari}
\affiliation{INO-CNR BEC Center and Dipartimento di Fisica, Universit\`a di Trento, 38123 Povo, Italy}
\affiliation{Trento Institute for Fundamental Physics and Applications, INFN, 38123 Povo, Italy}

\date{\today}

\begin{abstract}
	
We report on the measurement of the spin-dipole (SD) polarizability and of the frequency of the SD oscillation of a two-component Bose--Einstein condensate of sodium atoms  occupying the $|3^2S_{1/2}, F=1, m_F=\pm1\rangle$ hyperfine states. This binary spin-mixture presents the important properties of being, at the same time, fully miscible and rid of the limit set by buoyancy. It is also characterized by a huge enhancement of the SD polarizability and by the consequent softening of the frequency of the SD oscillation, due to the vicinity to the transition to the immiscible phase. The experimental data are successfully compared with the predictions of theory.
\end{abstract}

\maketitle


\section{Introduction}

The study of mixtures of Bose-Einstein condensates (BECs) has opened rich opportunities for novel experimental and theoretical investigations. Mixtures of ultracold atoms offer great flexibility thanks to the variety of atomic species and the additional degree of freedom related to the hyperfine structure \cite{Myatt97,Stenger98,Pu98,Ho98,Timmermans98,Ohmi98} (for a recent overview see \cite{Stamper13}). For a weakly interacting mixture of two BECs, the ground state of the system can either be a miscible mixture of the two components or a phase separated configuration \cite{Colson78}. Nevertheless, the stability of mixtures very close to the critical region is sensitive to other effects, such as asymmetries in the trapping potential \cite{Kevrekidis07}. Moreover, for systems in which the intracomponent coupling constants do not exactly coincide, one of the two components will experience a  positive buoyancy and will ``float'' on the other. Previous experiments involving two internal states of rubidium were affected by both of these problems \cite{Hall98,Weld10,Hamner11,Nicklas15,Eto16} hence setting strong limits to explore the many-body properties of miscible binary BECs. In particular, such conditions prevent the study of the static and dynamic response of an unpolarized system close to the transition between the miscible and immiscible phases, where interaction effects are particularly important despite the weakly interacting nature of the gas \cite{Jezek02}.

Here we report on the first measurement of the spin-dipole (SD) polarizability of a two-component BEC, as well as the frequency of the SD oscillation, by using an ultracold mixture of the $|3^2S_{1/2}, F=1, m_F=+1\rangle \equiv \ket{\uparrow}$ and $|3^2S_{1/2}, F=1, m_F=-1\rangle \equiv \ket{\downarrow}$ states of atomic sodium. The polarizability characterizes in a fundamental way the
thermodynamic behavior of binary ultracold gases and exhibits a divergent behavior at the transition between the miscible and immiscible phases, with the occurrence of important spin
fluctuations \cite{Recati11,Abad13,Bisset15}. On the other hand, the SD oscillation is the simplest collective excitation supported by the system in the presence of harmonic trapping and is characterized by the motion of the two components with opposite phase around equilibrium. The SD oscillation is  the analog of the famous giant dipole resonance of nuclear physics, where neutrons and protons oscillate with opposite phase \cite{Bohr69}. Actually, collective modes are a popular subject of research in quantum many body systems (see, \emph{e.g.}, \cite{Pitaevskii16}) where experiments are able to determine the corresponding frequencies with high precision, providing a good testbed for detailed comparison with theory and an accurate determination of the relevant interaction parameters. Collective dynamics has been already investigated in quantum binary mixtures of atomic gases like repulsive gases of Fermi atoms \cite{Vichi99,DeMarco02,Recati11,Valtolina16}, Bose-Bose \cite{Hall98,Sinatra99,Maddaloni00,Modugno01,Modugno02,Jezek02,Mertes07,Nicklas11,Zhang12,Egorov13,Sartori13,Sartori15} and Bose-Fermi mixtures \cite{Ferlaino03} as well as Bose-Fermi superfluid mixtures \cite{Ferrier14,Roy16}.
In the case of Bose-Bose mixtures both the polarization and the SD oscillation frequency are predicted to be crucially sensitive to the difference between the value of the intra and intercomponent interactions \cite{Jezek02,Sartori15} which is particularly small in our case. The dramatic change of the density profile of the trapped gas, caused by a small displacement of the minima of the trapping potentials of the two species near the miscible-immiscible phase transition, was first investigated theoretically in \cite{Jezek02}. 
Our mixture is not subject to buoyancy as $g_{\uparrow\uparrow}=g_{\downarrow\downarrow}\equiv g$ and is on the miscible side $g_{\uparrow\downarrow} < g$ near the boundary of the phase transition ($g$ and $g_{\uparrow\downarrow}$ are respectively the intra and intercomponent coupling constants). The  fact that  $(g - g_{\uparrow\downarrow}) / g \simeq 7 \, \%$, as given by the scattering lengths $a_{\uparrow\uparrow}=a_{\downarrow\downarrow}=54.54(20) a_0$ and $a_{\uparrow\downarrow}=50.78(40)a_0$, where $a_0$ is the Bohr radius \cite{Knoop11}, ensures the  stability of the mixture and, together with the absence of buoyancy, allows us to overcome the ultimate limits to measure both the polarizability and SD oscillation frequency.


\section{Mixture preparation}

\begin{figure}[t]
\centerline{{\includegraphics[width=0.5\textwidth]{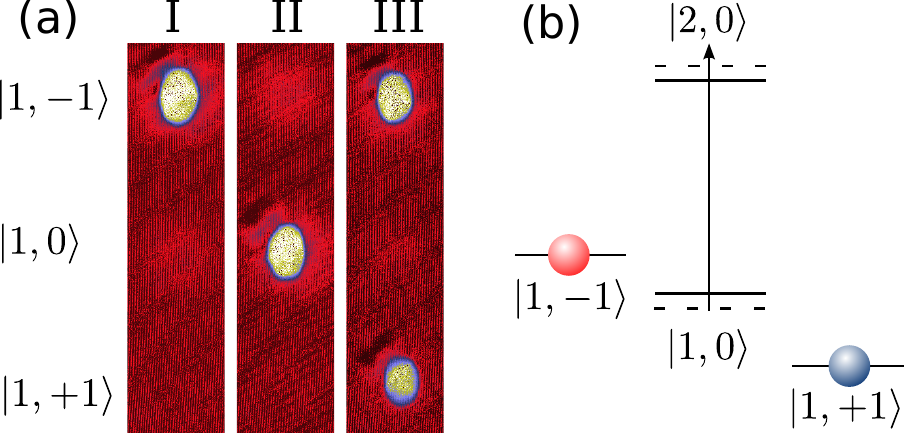}}}
\caption{(a) Absorption images taken after a SG expansion for I) the dipole loading, II) the Landau--Zener transition, III) the Rabi pulse leading to the creation of the binary mixture. (b) Stabilization of the two components by shifting the $|0\rangle$ state using microwave dressing on the transition to  $|F=2, m_F=0\rangle$.  }
\label{Fig1}
\end{figure}

Our experiment is based on the apparatus introduced in \cite{Lamporesi13} and starts with a nearly pure BEC of $^{23}$Na atoms in the $\ket{\downarrow}$ state in a crossed optical dipole trap with frequencies $\left[ \omega_x,\omega_y,\omega_z \right] / 2 \pi = \left[47.7(2),207.2(3),156.8(2)  \right] \, \text{Hz}$. The magnetic fields along the three spatial directions are calibrated with a precision of $1 \, \text{mG}$ using RF spectroscopy techniques.
The first step towards the creation of the spin mixture is to perform a Landau--Zener transition to the $|F=1, m_F=0\rangle \equiv \ket 0
$ state with nearly $100 \, \%$ transfer efficiency.
This is realized at a magnetic field of $100 \, \text{G}$ to isolate a two-level system exploiting the quadratic Zeeman shifts. The second step consists in inducing a Rabi oscillation among the three Zeeman sublevels to obtain a 50/50 spin mixture of $\ket{\downarrow}$ and $\ket{\uparrow}$ \cite{Zibold16}. 
The bias field along $\hat x$ is taken small enough to allow us to neglect the quadratic Zeeman shifts compared to the Rabi frequency and is kept on during the whole experimental sequence following the Rabi pulse. The number of atoms in each spin component is $N_\uparrow = N_\downarrow \simeq 10^6$ and the total chemical potential of the cloud is $\mu_{\text{tot}}/k_{\text{B}} \simeq 200 \, \text{nK}$. Fig. \ref{Fig1}(a) shows typical absorption images of the spinor BEC after a $10 \, \text{ms}$ Stern--Gerlach (SG) expansion in a magnetic field  gradient along $\hat z$. In order to prevent the decay of the mixture to $|0\rangle$ by spin changing collisions, we lift this level by $\sim 10 \, \text{kHz}$ using blue detuned microwave dressing on the transition to $|F=2, m_F=0\rangle$ (see Fig. \ref{Fig1}(b)).

\section{Spin-dipole polarizability}

\begin{figure}[t]
	\centerline{{\includegraphics[width=0.5\textwidth]{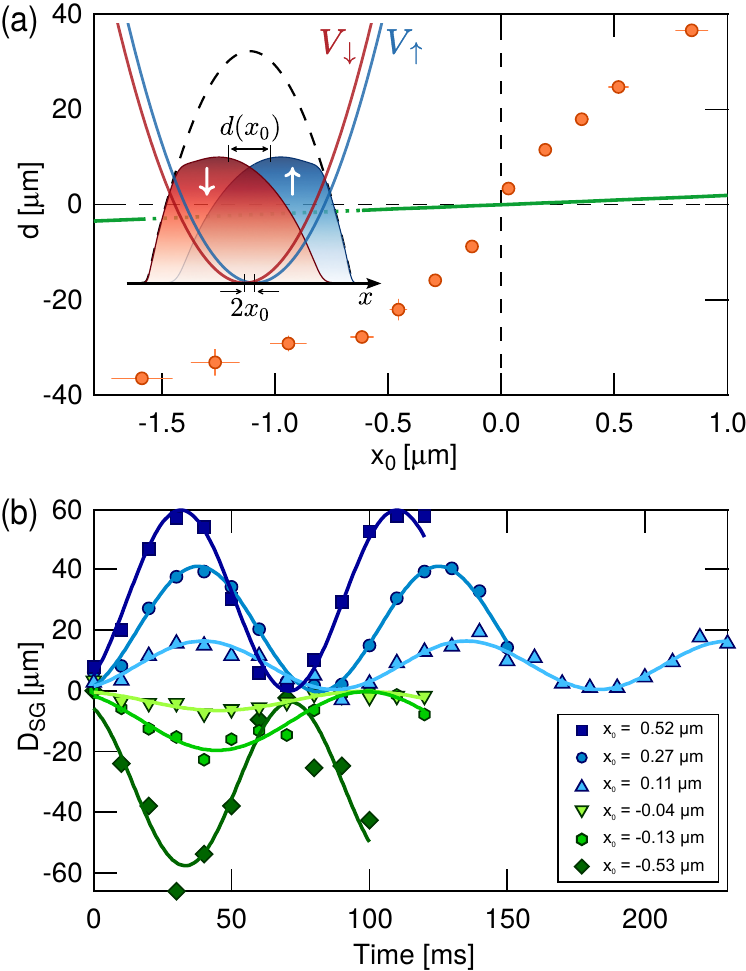}}}
	\caption{(a) Relative displacement $d =x_\downarrow-x_\uparrow$ between the spin components as a function of $x_0$ (orange dots). The green solid line corresponds to the situation of no intercomponent interaction $d = 2 x_0$. The figure also shows a sketch of the experimental conditions (the dashed curve is the total cloud density). (b) SD oscillations for different values of $x_0$ (positive and negative) observed using the second experimental protocol. The solid lines are fit to the data according to Eq. (\ref{SGEvol}). In each figure of the paper, data error bars are the sum in quadrature of systematic and statistical errors (one standard deviation of the mean).}
	\label{Fig2}
\end{figure}

The SD polarizability of a spin mixture describes the ability of the system to adapt itself to a displacement in opposite direction of the trapping potentials of the two components. After realizing a fully overlapped configuration, we adiabatically apply a magnetic field gradient $B'_x$ along $\hat x$ using a pair of coils in anti-Helmholtz configuration. The gradient is controlled with a resolution at the level of $4 \, \text{mG/cm}$. This displaces the minima of the trapping potentials such that $V_{\uparrow,\downarrow}=m\omega_x^2(x\pm x_0)^2/2$ where $ x_0 = g_F \mu_B B'_x/( m \omega_x^2)$ ($g_F$ is the Land\'e factor, $\mu_B$ the Bohr magneton and $m$ the atomic mass). The SD polarizability is defined as
\begin{equation}
\mathcal{P}(x_0) \equiv \frac{d(x_0)}{2 x_0},
\end{equation}
where $d=x_\downarrow-x_\uparrow$ is the \emph{in-situ} relative displacement between the centers of mass of each component  (see Fig. \ref{Fig2}(a)). After a $2 \, \text{ms}$ SG expansion, we measure $d$ by fitting each spin component density distribution to independent Thomas--Fermi (TF) profiles to extract their centers $x_{\uparrow,\downarrow}$. The individual density profiles are not exactly TF-like, but we verified, using a Gross--Pitaevskii equation (GPE) simulation, that this approximate fitting procedure results in an overestimation of  $\mathcal{P}$ by at most 6\%. Later in the text we discuss the additional correction to the measurement of $\mathcal{P}$ related to interactions during the SG expansion. Fig. \ref{Fig2}(a) shows the experimental results where the value of $x_0$ is estimated after calibrating $B'_x$. We observe that all data points strongly deviate from the prediction $d=2 x_0$ for a mixture without intercomponent interactions (green solid line), revealing the large SD polarizability of the system.

We use a second experimental protocol to determine the polarizability which will later prove to be useful for measuring the SD oscillation frequency. It consists in realizing the Rabi pulse $\ket 0 \rightarrow \ket{\uparrow,\downarrow}$ in the presence of a magnetic field gradient. As the minima of the trapping potentials for the $\ket{\uparrow}$ and $\ket{\downarrow}$ states are shifted by $\mp x_0$ with respect to the initial state $|0 \rangle$, this makes the two components oscillate out of phase after the Rabi pulse.  The \emph{in-situ} time evolution of the relative displacement $D(t)=x_\downarrow(t)-x_\uparrow(t)$ is expected to be given by $D(x_0,t) = d(x_0) \left[1- \cos \left[ \omega(x_0) t \right] \right]$. Measurements of such oscillations after a SG expansion of $t_{\text{SG}} = 10 \, \text{ms}$ for different values of $x_0$ varying the magnetic field gradient are reported in  Fig. \ref{Fig2}(b). After the SG expansion, the displacement between the spin components is given by $D_{\text{SG}}(x_0,t,t_{\text{SG}}) = D(x_0,t) + \partial_t D (x_0,t) \, t_{\text{SG}}$ such that we analyze the data by fitting it with the following function:
\begin{equation} \label{SGEvol}
D_{\text{SG}} = A(x_0,t_{\text{SG}}) \cos \left[ \omega(x_0) t + \phi(x_0,t_{\text{SG}})  \right] + d(x_0),
\end{equation}
where $A(x_0,t_{SG} ) = -d(x_0) \sqrt{1+\omega^2(x_0) t^2_{\text{SG}} } $ and $ \phi(x_0,t_{\text{SG}}) = \arctan \left[ \omega(x_0) t_{\text{SG}}\right]$. Eq. (\ref{SGEvol}) allows us to extract the value of $d(x_0)$ neglecting here again intercomponent interactions during the expansion.

\begin{figure}[t]
	\centerline{{\includegraphics[width=0.5\textwidth]{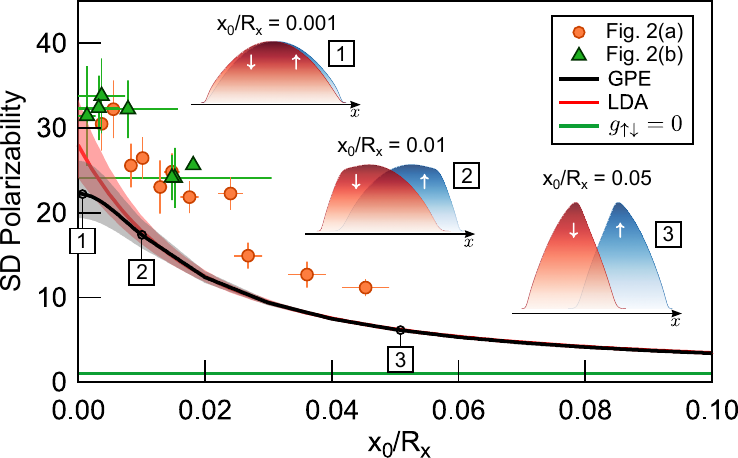}}}
	\caption{SD polarizability extracted from the data of Fig. \ref{Fig2}(a) (orange dots) and (b) (green triangles). The black (red) solid line is the prediction computed using the GPE (LDA). The shaded regions give the uncertainties taking into account error bars on the value of the coupling constants \cite{Knoop11}. The green solid line corresponds to the situation of no intercomponent interaction $\mathcal P = 1$. We also provide the density profiles $n_{\uparrow,\downarrow}(x,0,0)$ from the GPE for $x_0/R_x=0.001, 0.01, 0.05$. The experimental points overestimate the actual value of $\mathcal{P}$ due to the approximation of the TF fit and the interaction effect during the expansion (see text).}
	\label{Fig3}
\end{figure}

Fig. \ref{Fig3} shows the SD polarizability as a function of $x_0/R_x$ ($R_x$ is the TF radius along $\hat x$) using the data of Fig. \ref{Fig2}. We notice a strong nonlinear dependence of the polarizability on the separation between the two trapping potential minima, which is maximal in the linear limit ($x_0 \rightarrow 0$) and tends to $1$ for large separation ($x_0\gg R_x$). In the same figure, we also plot the theoretical predictions obtained within the local density approximation (LDA) and the numerical integration of the GPE performed with the experimental parameters. We identify three regions along $\hat x$, with the outer two regions occupied by either the $\ket{\uparrow}$ or $\ket{\downarrow}$ component and the inner region occupied by both of them. In the linear limit ($x_0 \to 0$) the LDA gives the result
\begin{equation}
\label{LDAP}
\mathcal{P}(x_0 \rightarrow 0) =  \frac{g+g_{\uparrow\downarrow}}{g-g_{\uparrow\downarrow}},
\end{equation}
for the polarizability \cite{Sartori15}\footnote{The SD polarizability Eq. (\ref{LDAP}) should not be confused with the magnetic polarizability $\chi_M = 1/\left[n(g-g_{\uparrow\downarrow})\right]$ which is defined in uniform matter in terms of the energy cost $\delta E = M^2/(2 \chi_M)$ associated with a small polarization $M=(N_\uparrow-N_\downarrow)/(N_\uparrow+N_\downarrow)$ of the gas.}, pointing out its divergent behavior near the phase transition occurring at $g_{\uparrow\downarrow}=g$. The agreement between the LDA and the GPE is excellent except in the region of small minima separation where the LDA becomes less and less adequate because of the large value of the spin healing length $\hbar/\sqrt{2mn(g-g_{\uparrow\downarrow})}$ ($n$ is the total density of the cloud).
In general, we observe a good agreement between the theoretical predictions and the experimental data. In particular, the huge effect on the polarizability caused by the vicinity to the miscible-immiscible phase transition is clearly revealed and the scaling with $x_0/R_x$ is well reproduced. The data analysis presented so far has however been performed neglecting interactions between the spin components during the SG expansion. Indeed, GPE simulations of the expansion in the presence of interactions show that the experimentally measured polarizability is overestimated by $5 \%$ ($30 \%$) for the $2 \, \text{ms}$ ($10 \, \text{ms}$) SG expansion. This explains the remaining difference between the experimental points of Fig. \ref{Fig3} and the theoretical predictions.


\section{Spin-dipole oscillation}

A useful estimate of the SD frequency is obtained by employing a sum rule approach \cite{Pitaevskii16} based on the ratio  $\hbar^2 \omega_{\text{SD}}^2 = M_1/M_{-1}$, where $M_1=N \hbar^2/2m$ ($N=N_\uparrow+N_\downarrow$) is the model independent energy weighted sum rule relative to the SD operator  $\sum_{i} (x_{i\downarrow} - x_{i\uparrow})$, and $M_{-1}=N\mathcal{P}(x_0 \rightarrow 0)/(2m\omega_x^2)$ is the inverse energy weighted sum rule fixed according to linear response theory by the linear SD polarizability  \cite{Sartori15}. This leads to the following relation between the SD frequency and polarizability
\begin{equation}   \label{relationPolSD}
\omega_{\text{SD}}= \frac{\omega_x}{\sqrt{\mathcal{P}(x_0 \rightarrow 0)}} \, .
\end{equation}
Using the LDA expression (\ref{LDAP}) for the polarizability one derives the following prediction for the SD frequency
\begin{equation} \label{SD_ratio}
\omega_{\text{SD}}= \sqrt{ \frac{g-g_{\uparrow\downarrow}}{g+g_{\uparrow\downarrow}}} \, \omega_x \, .
\end{equation}
The same result can be directly obtained by generalizing the hydrodynamic theory developed in \cite{Stringari96} for density oscillations to the case of SD oscillations \cite{Pitaevskii16}. Eq. (\ref{SD_ratio}) explicitly points out the crucial role played by the intercomponent coupling constant $g_{\uparrow\downarrow}$ in softening the frequency of the SD mode with respect to the  value $\omega_x$ characterizing the frequency of the  in-phase center-of-mass oscillation. We check, using  time-dependent GPE simulations of the SD oscillations for our experimental parameters, that the sum rule prediction (4) provides $\omega_{SD}$ with an accuracy better than $1\%$ when substituting the value of the static SD polarizability $\mathcal P(x_0\rightarrow0)$ from the GPE. This demonstrates that an accurate SD frequency measurement can  be used to determine the value of the SD polarizability.

\begin{figure}[t]
	\centerline{{\includegraphics[width=0.5\textwidth]{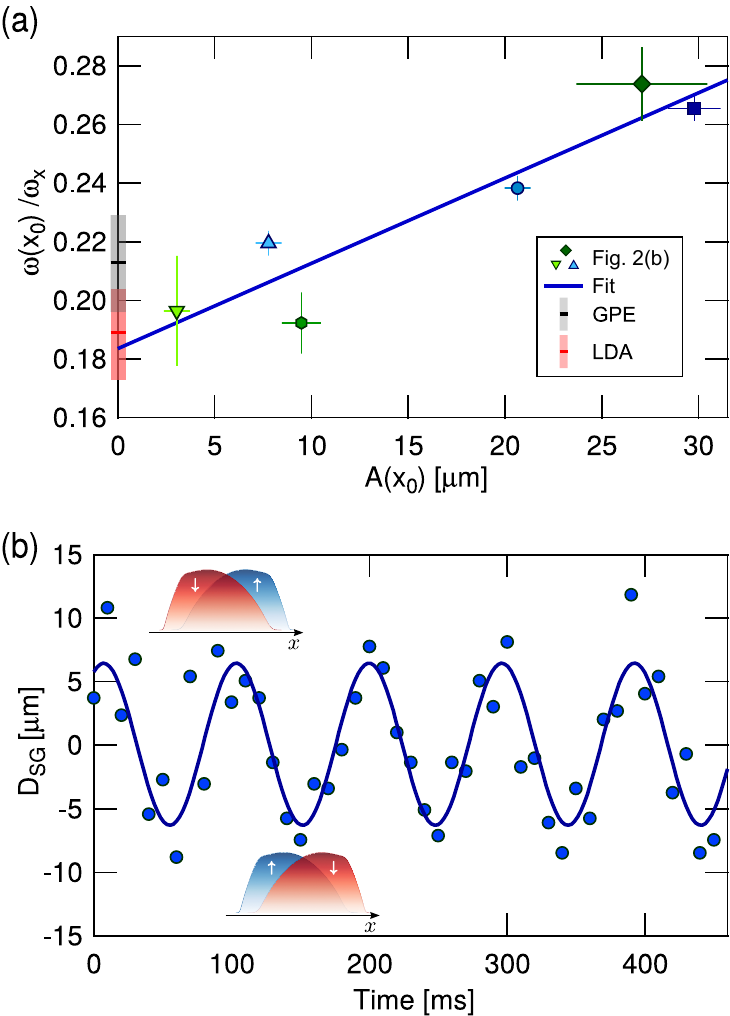}}}
	\caption{(a) Ratio $\omega(x_0)/\omega_x$ as a function of the amplitude $A(x_0)$ for the data of Fig. \ref{Fig2}(b) (same marker styles).  The black (red) marker is the prediction of the GPE $\omega_{\text{SD}}/\omega_x = 0.213(17)$ (LDA $\omega_{\text{SD}}/\omega_x = 0.189(15)$). Extrapolating the linear fit of $\omega(x_0)/\omega_x$  (solid blue line) for vanishing amplitude gives $\omega_{\text{SD}} /\omega_x = 0.18(1)$. (b) SD oscillations using the alternative method (blue dots) giving  $\omega_{\text{SD}}/\omega_x = 0.218(2)$. We also show two density profiles $n_{\uparrow,\downarrow}(x,0,0)$ illustrating the out-of-phase SD oscillations (data obtained from the GPE taking the equilibrium state for $x_0/R_x=0.01$ as initial condition before setting $x_0=0$ to start the oscillations).}
	\label{Fig4}
\end{figure}

As shown on Fig. \ref{Fig2}(b), the Rabi pulse in the presence of a magnetic field gradient gives rise to the excitation of SD oscillations whose frequency can be extracted as a function of the induced displacement $x_0$. A first estimate of the SD frequency is obtained considering that $\omega_{\text{SD}} = \omega(x_0 \rightarrow 0)$. Indeed, for large values of $x_0$, the oscillation frequency $\omega(x_0)$ approaches  $\omega_x$ while it decreases to $\omega_{\text{SD}}$ as $x_0 \rightarrow 0$. Since in the small $x_0$ limit the amplitude of the oscillation tends to zero, we  perform a linear fit to the curve of $\omega(x_0)/\omega_x$ as a function of the oscillation amplitude $A(x_0)$ and extract $\omega_{\text{SD}}/\omega_x=0.18(1)$ from the y-intercept of the linear fit (see Fig. \ref{Fig4}(a)). This method shows a good agreement with the LDA prediction Eq. (\ref{SD_ratio}) $\omega_{\text{SD}}/\omega_x = 0.189(15)$ and with the GPE simulations yielding $\omega_{\text{SD}}/\omega_x = 0.213(17)$ (uncertainties take into account error bars on the value of the coupling constants \cite{Knoop11}). The different values of $\omega_{\text{SD}}$ from the LDA and GPE calculations have the same origin as the one discussed in the case of the polarizability and are due to the large value of the spin healing length in the vicinity of the quantum phase transition.

An alternative and more efficient way to excite the SD mode and to measure its frequency consists in first creating two perfectly overlapped spin states where $B'_x = 0$ ($x_0=0$) and then applying a magnetic field gradient $B'_x = 0.1 \, \text{G/cm}$ ($x_0 = 1.3 \, \mu \text{m}$) for $3 \, \text{ms}  \ll 2 \pi / \omega(x_0)$ before restoring $B'_x = 0$. This leads to an \emph{in-situ} dipole oscillation shown in Fig. \ref{Fig4}(b) after $10 \, \text{ms}$ of SG expansion. We measure $\omega_{\text{SD}}/\omega_x = 0.218(2)$ which is slightly larger than the previous estimate based on the data of Fig. \ref{Fig2}(b) and shows better agreement with the prediction from the GPE simulations. For a precise determination of $\omega_{\text{SD}}$, it is important to ensure that the SD mode has a small \emph{in-situ} amplitude: here we estimate $D_{\text{SD}} = D_{\text{SG}} / \sqrt{1+\omega^2_{\text{SD}} t^2_{\text{SG}}} =  5.4 \, \mu\text{m}$ which is relatively small compared to the TF radius $R_x=40 \, \mu\text{m}$. In all experiments, we observe oscillations without noticeable damping on very long timescales (they are ultimately limited by the cloud lifetime). Indeed, the maximal relative velocities of the two superfluid components $v_{\text{max}} = 1.2 \, \text{mm/s}$  for the data of Fig. 2(b), and $v_{\text{max}} = 0.4 \, \text{mm/s}$  for the data of Fig. 4(b) are smaller than the critical velocity for the dynamical counterflow instability $v_{\text{cr}} =  \sqrt{\mu_{\text{tot}}(1- g_{\uparrow\downarrow}/g)/2m} = 1.8 \, \text{mm/s}$ \cite{Abad15}.


\section{Conclusion}

In conclusion, we reported on the experimental measurements of the polarizability and of the frequency of the SD oscillation in a two-component BEC of sodium. Because of the vicinity to the miscible-immiscible quantum phase transition both quantities are very sensitive to the value of the intercomponent interaction and their behavior deviates by large factors from the  values predicted in the absence of intercomponent interaction. This represents a major difference with respect to other available superfluid quantum mixtures, like the Bose-Fermi mixtures of lithium gases \cite{Ferrier14,Delehaye15}, where the role played by the intercomponent interaction is much less crucial. Similarly to the case of \cite{Ferrier14,Delehaye15} our mixture is characterized by  two interacting superfluids oscillating with opposite phase and the observed SD oscillation is undamped for small amplitude as a consequence of superfluidity. For large amplitude motion the  Landau's critical velocity will, however, behave very differently, being very  sensitive to the value of the intercomponent interaction \cite{Abad15}. Another interesting feature concerns the behavior of the SD oscillation at finite temperature. While the damping of the SD oscillation was actually observed in the old experiments of \cite{DeMarco02} carried out on a normal Fermi gas, understanding the behavior of the collective modes in the presence of both a condensed  (superfluid) and thermal  (non-superfluid) components remains extremely challenging \cite{Armaitis15, Lee16}. Other topics of interest concern the experimental realization  of magnetic solitons \cite{Qu16} and the inclusion of   coherent coupling between the two spin components. The Bose mixtures realized and investigated here then represent an ideal platform to explore important equilibrium and dynamic properties of binary superfluids.

\begin{acknowledgments}
    We thank C. Salomon and G. Roati for discussions and critical reading of the manuscript, and L. Festa for technical assistance at the early stage of the experiment. We acknowledge funding by  the  Provincia  Autonoma  di  Trento, the QUIC grant of the Horizon 2020 FET program, and by the Istituto Nazionale di Fisica Nucleare.
\end{acknowledgments}


\begin{thebibliography}{45}
\expandafter\ifx\csname natexlab\endcsname\relax\def\natexlab#1{#1}\fi
\expandafter\ifx\csname bibnamefont\endcsname\relax
  \def\bibnamefont#1{#1}\fi
\expandafter\ifx\csname bibfnamefont\endcsname\relax
  \def\bibfnamefont#1{#1}\fi
\expandafter\ifx\csname citenamefont\endcsname\relax
  \def\citenamefont#1{#1}\fi
\expandafter\ifx\csname url\endcsname\relax
  \def\url#1{\texttt{#1}}\fi
\expandafter\ifx\csname urlprefix\endcsname\relax\def\urlprefix{URL }\fi
\providecommand{\bibinfo}[2]{#2}
\providecommand{\eprint}[2][]{\url{#2}}

\bibitem[{\citenamefont{Myatt et~al.}(1997)\citenamefont{Myatt, Burt, Ghrist,
  Cornell, and Wieman}}]{Myatt97}
\bibinfo{author}{\bibfnamefont{C.~J.} \bibnamefont{Myatt}},
  \bibinfo{author}{\bibfnamefont{E.~A.} \bibnamefont{Burt}},
  \bibinfo{author}{\bibfnamefont{R.~W.} \bibnamefont{Ghrist}},
  \bibinfo{author}{\bibfnamefont{E.~A.} \bibnamefont{Cornell}},
  \bibnamefont{and} \bibinfo{author}{\bibfnamefont{C.~E.}
  \bibnamefont{Wieman}}, \bibinfo{journal}{Phys. Rev. Lett.}
  \textbf{\bibinfo{volume}{78}}, \bibinfo{pages}{586} (\bibinfo{year}{1997}).

\bibitem[{\citenamefont{Stenger et~al.}(1998)\citenamefont{Stenger, Inouye,
  Stamper-Kurn, Miesner, Chikkatur, and Ketterle}}]{Stenger98}
\bibinfo{author}{\bibfnamefont{J.}~\bibnamefont{Stenger}},
  \bibinfo{author}{\bibfnamefont{S.}~\bibnamefont{Inouye}},
  \bibinfo{author}{\bibfnamefont{D.~M.} \bibnamefont{Stamper-Kurn}},
  \bibinfo{author}{\bibfnamefont{H.-J.} \bibnamefont{Miesner}},
  \bibinfo{author}{\bibfnamefont{A.~P.} \bibnamefont{Chikkatur}},
  \bibnamefont{and} \bibinfo{author}{\bibfnamefont{W.}~\bibnamefont{Ketterle}},
  \bibinfo{journal}{Nature} \textbf{\bibinfo{volume}{396}},
  \bibinfo{pages}{345} (\bibinfo{year}{1998}).

\bibitem[{\citenamefont{Pu and Bigelow}(1998)}]{Pu98}
\bibinfo{author}{\bibfnamefont{H.}~\bibnamefont{Pu}} \bibnamefont{and}
  \bibinfo{author}{\bibfnamefont{N.~P.} \bibnamefont{Bigelow}},
  \bibinfo{journal}{Phys. Rev. Lett.} \textbf{\bibinfo{volume}{80}},
  \bibinfo{pages}{1130} (\bibinfo{year}{1998}).

\bibitem[{\citenamefont{Ho}(1998)}]{Ho98}
\bibinfo{author}{\bibfnamefont{T.-L.} \bibnamefont{Ho}},
  \bibinfo{journal}{Phys. Rev. Lett.} \textbf{\bibinfo{volume}{81}},
  \bibinfo{pages}{742} (\bibinfo{year}{1998}).

\bibitem[{\citenamefont{Timmermans}(1998)}]{Timmermans98}
\bibinfo{author}{\bibfnamefont{E.}~\bibnamefont{Timmermans}},
  \bibinfo{journal}{Phys. Rev. Lett.} \textbf{\bibinfo{volume}{81}},
  \bibinfo{pages}{5718} (\bibinfo{year}{1998}).

\bibitem[{\citenamefont{Ohmi and Machida}(1998)}]{Ohmi98}
\bibinfo{author}{\bibfnamefont{T.}~\bibnamefont{Ohmi}} \bibnamefont{and}
  \bibinfo{author}{\bibfnamefont{K.}~\bibnamefont{Machida}},
  \bibinfo{journal}{Journal of the Physical Society of Japan}
  \textbf{\bibinfo{volume}{67}}, \bibinfo{pages}{1822} (\bibinfo{year}{1998}).

\bibitem[{\citenamefont{Stamper-Kurn and Ueda}(2013)}]{Stamper13}
\bibinfo{author}{\bibfnamefont{D.~M.} \bibnamefont{Stamper-Kurn}}
  \bibnamefont{and} \bibinfo{author}{\bibfnamefont{M.}~\bibnamefont{Ueda}},
  \bibinfo{journal}{Rev. Mod. Phys.} \textbf{\bibinfo{volume}{85}},
  \bibinfo{pages}{1191} (\bibinfo{year}{2013}).

\bibitem[{\citenamefont{Colson and Fetter}(1978)}]{Colson78}
\bibinfo{author}{\bibfnamefont{W.~B.} \bibnamefont{Colson}} \bibnamefont{and}
  \bibinfo{author}{\bibfnamefont{A.~L.} \bibnamefont{Fetter}},
  \bibinfo{journal}{Journal of Low Temperature Physics}
  \textbf{\bibinfo{volume}{33}}, \bibinfo{pages}{231} (\bibinfo{year}{1978}).

\bibitem[{\citenamefont{Kevrekidis et~al.}(2007)\citenamefont{Kevrekidis,
  Frantzeskakis, and Carretero-Gonz{\'a}lez}}]{Kevrekidis07}
\bibinfo{author}{\bibfnamefont{P.}~\bibnamefont{Kevrekidis}},
  \bibinfo{author}{\bibfnamefont{D.}~\bibnamefont{Frantzeskakis}},
  \bibnamefont{and}
  \bibinfo{author}{\bibfnamefont{R.}~\bibnamefont{Carretero-Gonz{\'a}lez}},
  \emph{\bibinfo{title}{Emergent Nonlinear Phenomena in Bose-Einstein
  Condensates: Theory and Experiment}}, Springer Series on Atomic, Optical, and
  Plasma Physics (\bibinfo{publisher}{Springer Berlin Heidelberg},
  \bibinfo{year}{2007}), ISBN \bibinfo{isbn}{9783540735915}.

\bibitem[{\citenamefont{Hall et~al.}(1998)\citenamefont{Hall, Matthews, Ensher,
  Wieman, and Cornell}}]{Hall98}
\bibinfo{author}{\bibfnamefont{D.~S.} \bibnamefont{Hall}},
  \bibinfo{author}{\bibfnamefont{M.~R.} \bibnamefont{Matthews}},
  \bibinfo{author}{\bibfnamefont{J.~R.} \bibnamefont{Ensher}},
  \bibinfo{author}{\bibfnamefont{C.~E.} \bibnamefont{Wieman}},
  \bibnamefont{and} \bibinfo{author}{\bibfnamefont{E.~A.}
  \bibnamefont{Cornell}}, \bibinfo{journal}{Phys. Rev. Lett.}
  \textbf{\bibinfo{volume}{81}}, \bibinfo{pages}{1539} (\bibinfo{year}{1998}).

\bibitem[{\citenamefont{Weld et~al.}(2010)\citenamefont{Weld, Miyake, Medley,
  Pritchard, and Ketterle}}]{Weld10}
\bibinfo{author}{\bibfnamefont{D.~M.} \bibnamefont{Weld}},
  \bibinfo{author}{\bibfnamefont{H.}~\bibnamefont{Miyake}},
  \bibinfo{author}{\bibfnamefont{P.}~\bibnamefont{Medley}},
  \bibinfo{author}{\bibfnamefont{D.~E.} \bibnamefont{Pritchard}},
  \bibnamefont{and} \bibinfo{author}{\bibfnamefont{W.}~\bibnamefont{Ketterle}},
  \bibinfo{journal}{Phys. Rev. A} \textbf{\bibinfo{volume}{82}},
  \bibinfo{pages}{051603} (\bibinfo{year}{2010}).

\bibitem[{\citenamefont{Hamner et~al.}(2011)\citenamefont{Hamner, Chang,
  Engels, and Hoefer}}]{Hamner11}
\bibinfo{author}{\bibfnamefont{C.}~\bibnamefont{Hamner}},
  \bibinfo{author}{\bibfnamefont{J.~J.} \bibnamefont{Chang}},
  \bibinfo{author}{\bibfnamefont{P.}~\bibnamefont{Engels}}, \bibnamefont{and}
  \bibinfo{author}{\bibfnamefont{M.~A.} \bibnamefont{Hoefer}},
  \bibinfo{journal}{Phys. Rev. Lett.} \textbf{\bibinfo{volume}{106}},
  \bibinfo{pages}{065302} (\bibinfo{year}{2011}).

\bibitem[{\citenamefont{Nicklas et~al.}(2015)\citenamefont{Nicklas, Muessel,
  Strobel, Kevrekidis, and Oberthaler}}]{Nicklas15}
\bibinfo{author}{\bibfnamefont{E.}~\bibnamefont{Nicklas}},
  \bibinfo{author}{\bibfnamefont{W.}~\bibnamefont{Muessel}},
  \bibinfo{author}{\bibfnamefont{H.}~\bibnamefont{Strobel}},
  \bibinfo{author}{\bibfnamefont{P.~G.} \bibnamefont{Kevrekidis}},
  \bibnamefont{and} \bibinfo{author}{\bibfnamefont{M.~K.}
  \bibnamefont{Oberthaler}}, \bibinfo{journal}{Phys. Rev. A}
  \textbf{\bibinfo{volume}{92}}, \bibinfo{pages}{053614}
  (\bibinfo{year}{2015}).

\bibitem[{\citenamefont{Eto et~al.}(2016)\citenamefont{Eto, Takahashi, Kunimi,
  Saito, and Hirano}}]{Eto16}
\bibinfo{author}{\bibfnamefont{Y.}~\bibnamefont{Eto}},
  \bibinfo{author}{\bibfnamefont{M.}~\bibnamefont{Takahashi}},
  \bibinfo{author}{\bibfnamefont{M.}~\bibnamefont{Kunimi}},
  \bibinfo{author}{\bibfnamefont{H.}~\bibnamefont{Saito}}, \bibnamefont{and}
  \bibinfo{author}{\bibfnamefont{T.}~\bibnamefont{Hirano}},
  \bibinfo{journal}{New Journal of Physics} \textbf{\bibinfo{volume}{18}},
  \bibinfo{pages}{073029} (\bibinfo{year}{2016}).

\bibitem[{\citenamefont{Jezek and Capuzzi}(2002)}]{Jezek02}
\bibinfo{author}{\bibfnamefont{D.~M.} \bibnamefont{Jezek}} \bibnamefont{and}
  \bibinfo{author}{\bibfnamefont{P.}~\bibnamefont{Capuzzi}},
  \bibinfo{journal}{Phys. Rev. A} \textbf{\bibinfo{volume}{66}},
  \bibinfo{pages}{015602} (\bibinfo{year}{2002}).

\bibitem[{\citenamefont{Recati and Stringari}(2011)}]{Recati11}
\bibinfo{author}{\bibfnamefont{A.}~\bibnamefont{Recati}} \bibnamefont{and}
  \bibinfo{author}{\bibfnamefont{S.}~\bibnamefont{Stringari}},
  \bibinfo{journal}{Phys. Rev. Lett.} \textbf{\bibinfo{volume}{106}},
  \bibinfo{pages}{080402} (\bibinfo{year}{2011}).

\bibitem[{\citenamefont{Abad and Recati}(2013)}]{Abad13}
\bibinfo{author}{\bibfnamefont{M.}~\bibnamefont{Abad}} \bibnamefont{and}
  \bibinfo{author}{\bibfnamefont{A.}~\bibnamefont{Recati}},
  \bibinfo{journal}{Eur. Phys. J. D} \textbf{\bibinfo{volume}{67}},
  \bibinfo{pages}{148} (\bibinfo{year}{2013}).

\bibitem[{\citenamefont{Bisset et~al.}(2015)\citenamefont{Bisset, Wilson, and
  Ticknor}}]{Bisset15}
\bibinfo{author}{\bibfnamefont{R.~N.} \bibnamefont{Bisset}},
  \bibinfo{author}{\bibfnamefont{R.~M.} \bibnamefont{Wilson}},
  \bibnamefont{and} \bibinfo{author}{\bibfnamefont{C.}~\bibnamefont{Ticknor}},
  \bibinfo{journal}{Phys. Rev. A} \textbf{\bibinfo{volume}{91}},
  \bibinfo{pages}{053613} (\bibinfo{year}{2015}).

\bibitem[{\citenamefont{Bohr and Mottelson}(1969)}]{Bohr69}
\bibinfo{author}{\bibfnamefont{A.}~\bibnamefont{Bohr}} \bibnamefont{and}
  \bibinfo{author}{\bibfnamefont{B.~R.} \bibnamefont{Mottelson}},
  \emph{\bibinfo{title}{Nuclear structure}} (\bibinfo{publisher}{World
  Scientific Publishing}, \bibinfo{year}{1969}).

\bibitem[{\citenamefont{Pitaevskii and Stringari}(2016)}]{Pitaevskii16}
\bibinfo{author}{\bibfnamefont{L.}~\bibnamefont{Pitaevskii}} \bibnamefont{and}
  \bibinfo{author}{\bibfnamefont{S.}~\bibnamefont{Stringari}},
  \emph{\bibinfo{title}{Bose-Einstein condensation and superfluidity}}
  (\bibinfo{publisher}{Oxford University Press}, \bibinfo{year}{2016}).

\bibitem[{\citenamefont{Vichi and Stringari}(1999)}]{Vichi99}
\bibinfo{author}{\bibfnamefont{L.}~\bibnamefont{Vichi}} \bibnamefont{and}
  \bibinfo{author}{\bibfnamefont{S.}~\bibnamefont{Stringari}},
  \bibinfo{journal}{Phys. Rev. A} \textbf{\bibinfo{volume}{60}},
  \bibinfo{pages}{4734} (\bibinfo{year}{1999}).

\bibitem[{\citenamefont{DeMarco and Jin}(2002)}]{DeMarco02}
\bibinfo{author}{\bibfnamefont{B.}~\bibnamefont{DeMarco}} \bibnamefont{and}
  \bibinfo{author}{\bibfnamefont{D.~S.} \bibnamefont{Jin}},
  \bibinfo{journal}{Phys. Rev. Lett.} \textbf{\bibinfo{volume}{88}},
  \bibinfo{pages}{040405} (\bibinfo{year}{2002}).

\bibitem[{\citenamefont{{Valtolina} et~al.}(2016)\citenamefont{{Valtolina},
  {Scazza}, {Amico}, {Burchianti}, {Recati}, {Enss}, {Inguscio}, {Zaccanti},
  and {Roati}}}]{Valtolina16}
\bibinfo{author}{\bibfnamefont{G.}~\bibnamefont{{Valtolina}}},
  \bibinfo{author}{\bibfnamefont{F.}~\bibnamefont{{Scazza}}},
  \bibinfo{author}{\bibfnamefont{A.}~\bibnamefont{{Amico}}},
  \bibinfo{author}{\bibfnamefont{A.}~\bibnamefont{{Burchianti}}},
  \bibinfo{author}{\bibfnamefont{A.}~\bibnamefont{{Recati}}},
  \bibinfo{author}{\bibfnamefont{T.}~\bibnamefont{{Enss}}},
  \bibinfo{author}{\bibfnamefont{M.}~\bibnamefont{{Inguscio}}},
  \bibinfo{author}{\bibfnamefont{M.}~\bibnamefont{{Zaccanti}}},
  \bibnamefont{and} \bibinfo{author}{\bibfnamefont{G.}~\bibnamefont{{Roati}}},
  \bibinfo{journal}{ArXiv e-prints}  (\bibinfo{year}{2016}),
  \eprint{1605.07850}.

\bibitem[{\citenamefont{Sinatra et~al.}(1999)\citenamefont{Sinatra, Fedichev,
  Castin, Dalibard, and Shlyapnikov}}]{Sinatra99}
\bibinfo{author}{\bibfnamefont{A.}~\bibnamefont{Sinatra}},
  \bibinfo{author}{\bibfnamefont{P.~O.} \bibnamefont{Fedichev}},
  \bibinfo{author}{\bibfnamefont{Y.}~\bibnamefont{Castin}},
  \bibinfo{author}{\bibfnamefont{J.}~\bibnamefont{Dalibard}}, \bibnamefont{and}
  \bibinfo{author}{\bibfnamefont{G.~V.} \bibnamefont{Shlyapnikov}},
  \bibinfo{journal}{Phys. Rev. Lett.} \textbf{\bibinfo{volume}{82}},
  \bibinfo{pages}{251} (\bibinfo{year}{1999}).

\bibitem[{\citenamefont{Maddaloni et~al.}(2000)\citenamefont{Maddaloni,
  Modugno, Fort, Minardi, and Inguscio}}]{Maddaloni00}
\bibinfo{author}{\bibfnamefont{P.}~\bibnamefont{Maddaloni}},
  \bibinfo{author}{\bibfnamefont{M.}~\bibnamefont{Modugno}},
  \bibinfo{author}{\bibfnamefont{C.}~\bibnamefont{Fort}},
  \bibinfo{author}{\bibfnamefont{F.}~\bibnamefont{Minardi}}, \bibnamefont{and}
  \bibinfo{author}{\bibfnamefont{M.}~\bibnamefont{Inguscio}},
  \bibinfo{journal}{Phys. Rev. Lett.} \textbf{\bibinfo{volume}{85}},
  \bibinfo{pages}{2413} (\bibinfo{year}{2000}).

\bibitem[{\citenamefont{Modugno et~al.}(2001)\citenamefont{Modugno, Fort,
  Maddaloni, Minardi, and Inguscio}}]{Modugno01}
\bibinfo{author}{\bibfnamefont{M.}~\bibnamefont{Modugno}},
  \bibinfo{author}{\bibfnamefont{C.}~\bibnamefont{Fort}},
  \bibinfo{author}{\bibfnamefont{P.}~\bibnamefont{Maddaloni}},
  \bibinfo{author}{\bibfnamefont{F.}~\bibnamefont{Minardi}}, \bibnamefont{and}
  \bibinfo{author}{\bibfnamefont{M.}~\bibnamefont{Inguscio}},
  \bibinfo{journal}{Eur. Phys. J. D} \textbf{\bibinfo{volume}{17}},
  \bibinfo{pages}{345} (\bibinfo{year}{2001}).

\bibitem[{\citenamefont{Modugno et~al.}(2002)\citenamefont{Modugno, Modugno,
  Riboli, Roati, and Inguscio}}]{Modugno02}
\bibinfo{author}{\bibfnamefont{G.}~\bibnamefont{Modugno}},
  \bibinfo{author}{\bibfnamefont{M.}~\bibnamefont{Modugno}},
  \bibinfo{author}{\bibfnamefont{F.}~\bibnamefont{Riboli}},
  \bibinfo{author}{\bibfnamefont{G.}~\bibnamefont{Roati}}, \bibnamefont{and}
  \bibinfo{author}{\bibfnamefont{M.}~\bibnamefont{Inguscio}},
  \bibinfo{journal}{Phys. Rev. Lett.} \textbf{\bibinfo{volume}{89}},
  \bibinfo{pages}{190404} (\bibinfo{year}{2002}).

\bibitem[{\citenamefont{Mertes et~al.}(2007)\citenamefont{Mertes, Merrill,
  Carretero-Gonz\'alez, Frantzeskakis, Kevrekidis, and Hall}}]{Mertes07}
\bibinfo{author}{\bibfnamefont{K.~M.} \bibnamefont{Mertes}},
  \bibinfo{author}{\bibfnamefont{J.~W.} \bibnamefont{Merrill}},
  \bibinfo{author}{\bibfnamefont{R.}~\bibnamefont{Carretero-Gonz\'alez}},
  \bibinfo{author}{\bibfnamefont{D.~J.} \bibnamefont{Frantzeskakis}},
  \bibinfo{author}{\bibfnamefont{P.~G.} \bibnamefont{Kevrekidis}},
  \bibnamefont{and} \bibinfo{author}{\bibfnamefont{D.~S.} \bibnamefont{Hall}},
  \bibinfo{journal}{Phys. Rev. Lett.} \textbf{\bibinfo{volume}{99}},
  \bibinfo{pages}{190402} (\bibinfo{year}{2007}).

\bibitem[{\citenamefont{Nicklas et~al.}(2011)\citenamefont{Nicklas, Strobel,
  Zibold, Gross, Malomed, Kevrekidis, and Oberthaler}}]{Nicklas11}
\bibinfo{author}{\bibfnamefont{E.}~\bibnamefont{Nicklas}},
  \bibinfo{author}{\bibfnamefont{H.}~\bibnamefont{Strobel}},
  \bibinfo{author}{\bibfnamefont{T.}~\bibnamefont{Zibold}},
  \bibinfo{author}{\bibfnamefont{C.}~\bibnamefont{Gross}},
  \bibinfo{author}{\bibfnamefont{B.~A.} \bibnamefont{Malomed}},
  \bibinfo{author}{\bibfnamefont{P.~G.} \bibnamefont{Kevrekidis}},
  \bibnamefont{and} \bibinfo{author}{\bibfnamefont{M.~K.}
  \bibnamefont{Oberthaler}}, \bibinfo{journal}{Phys. Rev. Lett.}
  \textbf{\bibinfo{volume}{107}}, \bibinfo{pages}{193001}
  (\bibinfo{year}{2011}).

\bibitem[{\citenamefont{Zhang et~al.}(2012)\citenamefont{Zhang, Ji, Chen,
  Zhang, Du, Yan, Pan, Zhao, Deng, Zhai et~al.}}]{Zhang12}
\bibinfo{author}{\bibfnamefont{J.-Y.} \bibnamefont{Zhang}},
  \bibinfo{author}{\bibfnamefont{S.-C.} \bibnamefont{Ji}},
  \bibinfo{author}{\bibfnamefont{Z.}~\bibnamefont{Chen}},
  \bibinfo{author}{\bibfnamefont{L.}~\bibnamefont{Zhang}},
  \bibinfo{author}{\bibfnamefont{Z.-D.} \bibnamefont{Du}},
  \bibinfo{author}{\bibfnamefont{B.}~\bibnamefont{Yan}},
  \bibinfo{author}{\bibfnamefont{G.-S.} \bibnamefont{Pan}},
  \bibinfo{author}{\bibfnamefont{B.}~\bibnamefont{Zhao}},
  \bibinfo{author}{\bibfnamefont{Y.-J.} \bibnamefont{Deng}},
  \bibinfo{author}{\bibfnamefont{H.}~\bibnamefont{Zhai}}, \bibnamefont{et~al.},
  \bibinfo{journal}{Phys. Rev. Lett.} \textbf{\bibinfo{volume}{109}},
  \bibinfo{pages}{115301} (\bibinfo{year}{2012}).

\bibitem[{\citenamefont{Egorov et~al.}(2013)\citenamefont{Egorov, Opanchuk,
  Drummond, Hall, Hannaford, and Sidorov}}]{Egorov13}
\bibinfo{author}{\bibfnamefont{M.}~\bibnamefont{Egorov}},
  \bibinfo{author}{\bibfnamefont{B.}~\bibnamefont{Opanchuk}},
  \bibinfo{author}{\bibfnamefont{P.}~\bibnamefont{Drummond}},
  \bibinfo{author}{\bibfnamefont{B.~V.} \bibnamefont{Hall}},
  \bibinfo{author}{\bibfnamefont{P.}~\bibnamefont{Hannaford}},
  \bibnamefont{and} \bibinfo{author}{\bibfnamefont{A.~I.}
  \bibnamefont{Sidorov}}, \bibinfo{journal}{Phys. Rev. A}
  \textbf{\bibinfo{volume}{87}}, \bibinfo{pages}{053614}
  (\bibinfo{year}{2013}).

\bibitem[{\citenamefont{Sartori and Recati}(2013)}]{Sartori13}
\bibinfo{author}{\bibfnamefont{A.}~\bibnamefont{Sartori}} \bibnamefont{and}
  \bibinfo{author}{\bibfnamefont{A.}~\bibnamefont{Recati}},
  \bibinfo{journal}{Eur. Phys. J. D} \textbf{\bibinfo{volume}{67}},
  \bibinfo{pages}{260} (\bibinfo{year}{2013}).

\bibitem[{\citenamefont{Sartori et~al.}(2015)\citenamefont{Sartori, Marino,
  Stringari, and Recati}}]{Sartori15}
\bibinfo{author}{\bibfnamefont{A.}~\bibnamefont{Sartori}},
  \bibinfo{author}{\bibfnamefont{J.}~\bibnamefont{Marino}},
  \bibinfo{author}{\bibfnamefont{S.}~\bibnamefont{Stringari}},
  \bibnamefont{and} \bibinfo{author}{\bibfnamefont{A.}~\bibnamefont{Recati}},
  \bibinfo{journal}{New J. Phys.} \textbf{\bibinfo{volume}{17}},
  \bibinfo{pages}{093036} (\bibinfo{year}{2015}).

\bibitem[{\citenamefont{Ferlaino et~al.}(2003)\citenamefont{Ferlaino, Brecha,
  Hannaford, Riboli, Roati, Modugno, and Inguscio}}]{Ferlaino03}
\bibinfo{author}{\bibfnamefont{F.}~\bibnamefont{Ferlaino}},
  \bibinfo{author}{\bibfnamefont{R.~J.} \bibnamefont{Brecha}},
  \bibinfo{author}{\bibfnamefont{P.}~\bibnamefont{Hannaford}},
  \bibinfo{author}{\bibfnamefont{F.}~\bibnamefont{Riboli}},
  \bibinfo{author}{\bibfnamefont{G.}~\bibnamefont{Roati}},
  \bibinfo{author}{\bibfnamefont{G.}~\bibnamefont{Modugno}}, \bibnamefont{and}
  \bibinfo{author}{\bibfnamefont{M.}~\bibnamefont{Inguscio}},
  \bibinfo{journal}{Journal of Optics B: Quantum and Semiclassical Optics}
  \textbf{\bibinfo{volume}{5}}, \bibinfo{pages}{S3} (\bibinfo{year}{2003}).

\bibitem[{\citenamefont{Ferrier-Barbut
  et~al.}(2014)\citenamefont{Ferrier-Barbut, Delehaye, Laurent, Grier, Pierce,
  Rem, Chevy, and Salomon}}]{Ferrier14}
\bibinfo{author}{\bibfnamefont{I.}~\bibnamefont{Ferrier-Barbut}},
  \bibinfo{author}{\bibfnamefont{M.}~\bibnamefont{Delehaye}},
  \bibinfo{author}{\bibfnamefont{S.}~\bibnamefont{Laurent}},
  \bibinfo{author}{\bibfnamefont{A.~T.} \bibnamefont{Grier}},
  \bibinfo{author}{\bibfnamefont{M.}~\bibnamefont{Pierce}},
  \bibinfo{author}{\bibfnamefont{B.~S.} \bibnamefont{Rem}},
  \bibinfo{author}{\bibfnamefont{F.}~\bibnamefont{Chevy}}, \bibnamefont{and}
  \bibinfo{author}{\bibfnamefont{C.}~\bibnamefont{Salomon}},
  \bibinfo{journal}{Science} \textbf{\bibinfo{volume}{345}},
  \bibinfo{pages}{1035} (\bibinfo{year}{2014}).

\bibitem[{\citenamefont{{Roy} et~al.}(2016)\citenamefont{{Roy}, {Green},
  {Bowler}, and {Gupta}}}]{Roy16}
\bibinfo{author}{\bibfnamefont{R.}~\bibnamefont{{Roy}}},
  \bibinfo{author}{\bibfnamefont{A.}~\bibnamefont{{Green}}},
  \bibinfo{author}{\bibfnamefont{R.}~\bibnamefont{{Bowler}}}, \bibnamefont{and}
  \bibinfo{author}{\bibfnamefont{S.}~\bibnamefont{{Gupta}}},
  \bibinfo{journal}{ArXiv e-prints}  (\bibinfo{year}{2016}),
  \eprint{1607.03221}.

\bibitem[{\citenamefont{Knoop et~al.}(2011)\citenamefont{Knoop, Schuster,
  Scelle, Trautmann, Appmeier, Oberthaler, Tiesinga, and Tiemann}}]{Knoop11}
\bibinfo{author}{\bibfnamefont{S.}~\bibnamefont{Knoop}},
  \bibinfo{author}{\bibfnamefont{T.}~\bibnamefont{Schuster}},
  \bibinfo{author}{\bibfnamefont{R.}~\bibnamefont{Scelle}},
  \bibinfo{author}{\bibfnamefont{A.}~\bibnamefont{Trautmann}},
  \bibinfo{author}{\bibfnamefont{J.}~\bibnamefont{Appmeier}},
  \bibinfo{author}{\bibfnamefont{M.~K.} \bibnamefont{Oberthaler}},
  \bibinfo{author}{\bibfnamefont{E.}~\bibnamefont{Tiesinga}}, \bibnamefont{and}
  \bibinfo{author}{\bibfnamefont{E.}~\bibnamefont{Tiemann}},
  \bibinfo{journal}{Phys. Rev. A} \textbf{\bibinfo{volume}{83}},
  \bibinfo{pages}{042704} (\bibinfo{year}{2011}).

\bibitem[{\citenamefont{Lamporesi et~al.}(2013)\citenamefont{Lamporesi,
  Donadello, Serafini, and Ferrari}}]{Lamporesi13}
\bibinfo{author}{\bibfnamefont{G.}~\bibnamefont{Lamporesi}},
  \bibinfo{author}{\bibfnamefont{S.}~\bibnamefont{Donadello}},
  \bibinfo{author}{\bibfnamefont{S.}~\bibnamefont{Serafini}}, \bibnamefont{and}
  \bibinfo{author}{\bibfnamefont{G.}~\bibnamefont{Ferrari}},
  \bibinfo{journal}{Rev. Sci. Instrum.} \textbf{\bibinfo{volume}{84}},
  \bibinfo{eid}{063102} (\bibinfo{year}{2013}).

\bibitem[{\citenamefont{Zibold et~al.}(2016)\citenamefont{Zibold, Corre,
  Frapolli, Invernizzi, Dalibard, and Gerbier}}]{Zibold16}
\bibinfo{author}{\bibfnamefont{T.}~\bibnamefont{Zibold}},
  \bibinfo{author}{\bibfnamefont{V.}~\bibnamefont{Corre}},
  \bibinfo{author}{\bibfnamefont{C.}~\bibnamefont{Frapolli}},
  \bibinfo{author}{\bibfnamefont{A.}~\bibnamefont{Invernizzi}},
  \bibinfo{author}{\bibfnamefont{J.}~\bibnamefont{Dalibard}}, \bibnamefont{and}
  \bibinfo{author}{\bibfnamefont{F.}~\bibnamefont{Gerbier}},
  \bibinfo{journal}{Phys. Rev. A} \textbf{\bibinfo{volume}{93}},
  \bibinfo{pages}{023614} (\bibinfo{year}{2016}).

\bibitem[{\citenamefont{Stringari}(1996)}]{Stringari96}
\bibinfo{author}{\bibfnamefont{S.}~\bibnamefont{Stringari}},
  \bibinfo{journal}{Phys. Rev. Lett.} \textbf{\bibinfo{volume}{77}},
  \bibinfo{pages}{2360} (\bibinfo{year}{1996}).

\bibitem[{\citenamefont{Abad et~al.}(2015)\citenamefont{Abad, Recati,
  Stringari, and Chevy}}]{Abad15}
\bibinfo{author}{\bibfnamefont{M.}~\bibnamefont{Abad}},
  \bibinfo{author}{\bibfnamefont{A.}~\bibnamefont{Recati}},
  \bibinfo{author}{\bibfnamefont{S.}~\bibnamefont{Stringari}},
  \bibnamefont{and} \bibinfo{author}{\bibfnamefont{F.}~\bibnamefont{Chevy}},
  \bibinfo{journal}{Eur. Phys. J. D} \textbf{\bibinfo{volume}{69}},
  \bibinfo{pages}{126} (\bibinfo{year}{2015}).

\bibitem[{\citenamefont{Delehaye et~al.}(2015)\citenamefont{Delehaye, Laurent,
  Ferrier-Barbut, Jin, Chevy, and Salomon}}]{Delehaye15}
\bibinfo{author}{\bibfnamefont{M.}~\bibnamefont{Delehaye}},
  \bibinfo{author}{\bibfnamefont{S.}~\bibnamefont{Laurent}},
  \bibinfo{author}{\bibfnamefont{I.}~\bibnamefont{Ferrier-Barbut}},
  \bibinfo{author}{\bibfnamefont{S.}~\bibnamefont{Jin}},
  \bibinfo{author}{\bibfnamefont{F.}~\bibnamefont{Chevy}}, \bibnamefont{and}
  \bibinfo{author}{\bibfnamefont{C.}~\bibnamefont{Salomon}},
  \bibinfo{journal}{Phys. Rev. Lett.} \textbf{\bibinfo{volume}{115}},
  \bibinfo{pages}{265303} (\bibinfo{year}{2015}).

\bibitem[{\citenamefont{Armaitis et~al.}(2015)\citenamefont{Armaitis, Stoof,
  and Duine}}]{Armaitis15}
\bibinfo{author}{\bibfnamefont{J.}~\bibnamefont{Armaitis}},
  \bibinfo{author}{\bibfnamefont{H.~T.~C.} \bibnamefont{Stoof}},
  \bibnamefont{and} \bibinfo{author}{\bibfnamefont{R.~A.} \bibnamefont{Duine}},
  \bibinfo{journal}{Phys. Rev. A} \textbf{\bibinfo{volume}{91}},
  \bibinfo{pages}{043641} (\bibinfo{year}{2015}).

\bibitem[{\citenamefont{Lee et~al.}(2016)\citenamefont{Lee, J\o{}rgensen, Liu,
  Wacker, Arlt, and Proukakis}}]{Lee16}
\bibinfo{author}{\bibfnamefont{K.~L.} \bibnamefont{Lee}},
  \bibinfo{author}{\bibfnamefont{N.~B.} \bibnamefont{J\o{}rgensen}},
  \bibinfo{author}{\bibfnamefont{I.-K.} \bibnamefont{Liu}},
  \bibinfo{author}{\bibfnamefont{L.}~\bibnamefont{Wacker}},
  \bibinfo{author}{\bibfnamefont{J.~J.} \bibnamefont{Arlt}}, \bibnamefont{and}
  \bibinfo{author}{\bibfnamefont{N.~P.} \bibnamefont{Proukakis}},
  \bibinfo{journal}{Phys. Rev. A} \textbf{\bibinfo{volume}{94}},
  \bibinfo{pages}{013602} (\bibinfo{year}{2016}).

\bibitem[{\citenamefont{Qu et~al.}(2016)\citenamefont{Qu, Pitaevskii, and
  Stringari}}]{Qu16}
\bibinfo{author}{\bibfnamefont{C.}~\bibnamefont{Qu}},
  \bibinfo{author}{\bibfnamefont{L.~P.} \bibnamefont{Pitaevskii}},
  \bibnamefont{and}
  \bibinfo{author}{\bibfnamefont{S.}~\bibnamefont{Stringari}},
  \bibinfo{journal}{Phys. Rev. Lett.} \textbf{\bibinfo{volume}{116}},
  \bibinfo{pages}{160402} (\bibinfo{year}{2016}).

\end{thebibliography}
\end{document}